\documentclass{emulateapj}
\usepackage{amsmath}
\usepackage{graphicx}
\usepackage{amsfonts}
\usepackage{natbib}
\usepackage{color}

\newcommand{\z}{{\it z}}

\begin{document}

\title[Quasi-Periodic modulation in PKS 0426-380]{Possible Quasi-Periodic modulation in the \z~=~1.1 $\gamma$-ray blazar PKS 0426-380}
\author{Peng-fei~Zhang\altaffilmark{1,3}, Da-hai~Yan\altaffilmark{2,4,5}, Neng-hui~Liao\altaffilmark{1}, Wei~Zeng\altaffilmark{3}, Jian-cheng~Wang\altaffilmark{2,4,5}, Li-Jia Cao\altaffilmark{3}}
\altaffiltext{1}{Key Laboratory of Dark Matter and Space Astronomy, Purple Mountain Observatory, Chinese Academy of Sciences, Nanjing 210008, China; zhangpengfee@pmo.ac.cn}
\altaffiltext{2}{Yunnan Observatory, Chinese Academy of Sciences, Kunming 650011, China; yandahai@ynao.ac.cn}
\altaffiltext{3}{Key Laboratory of Astroparticle Physics of Yunnan Province, Yunnan University, Kunming 650091, China}
\altaffiltext{4}{Center for Astronomical Mega-Science, Chinese Academy of Sciences, 20A Datun Road, Chaoyang District, Beijing, 100012, China}
\altaffiltext{5}{Key Laboratory for the Structure and Evolution of Celestial Objects, Chinese Academy of Sciences, Kunming 650011, China}

\begin{abstract}

We search for $\gamma$-ray and optical periodic modulations in a distant flat spectrum radio quasar (FSRQ) PKS 0426-380 (the redshift $z=1.1$).
Using two techniques (i.e., the maximum likelihood optimization and the exposure-weighted aperture photometry),
we obtain $\gamma$-ray light curves from \emph{Fermi}-LAT Pass 8 data covering from 2008 August to 2016 December.
We then analyze the light curves with the Lomb-Scargle Periodogram (LSP) and the Weighted Wavelet Z-transform (WWZ).
A $\gamma$-ray quasi-periodicity with a period of 3.35 $\pm$ 0.68 years is found at the significance-level of  $\simeq3.6\ \sigma$.
The optical-UV flux covering from 2005 August to 2013 April provided by ASI SCIENCE DATA CENTER is also analyzed, but no significant quasi-periodicity is found. It should be pointed out that the result of the optical-UV data could be tentative because of the incomplete of the data.
Further long-term multiwavelength monitoring of this FSRQ is needed to confirm its quasi-periodicity.

\end{abstract}

\bigskip
\bigskip

%%%%%
%Section 1 - Introduction
%%%%%

\section{INTRODUCTION}
\label{sec:intro}

Blazar is one type of active galactic nucleus (AGNs) who aim their jets almost
directly at Earth. Blazar emission is generally dominated by non-thermal radiation
over all frequencies ranging from radio to TeV $\gamma$ rays.
The typical multi-wavelength spectral energy distribution (SED) of a
blazar is characterized by two distinct bumps.
It is generally accepted that the first bump peaking in infrared to X-ray frequencies is
synchrotron emission from relativistic electrons in the jet.
The second bump peaking in MeV to GeV band could be produced via inverse Compton (IC) scattering of synchrotron photons \citep[i.e., synchrotron-self Compton: SSC; e.g.,][]{Maraschi1992,Tavecchio98,Finke08,Yan14} and external photons \citep[i.e., external Compton: EC; e.g.,][]{Dermer93,Sikora1994,Kang} by the same population of relativistic electrons that produce the synchrotron emission.

According to the emission lines' features, blazars are classified as
BL Lacertae objects (BL Lacs; having weak or no emission lines) and
flat spectrum radio quasars (FSRQs; having strong emission lines).
FSRQs are usually the low synchrotron-peaked blazars (i.e., the synchrotron peak frequency $\nu_{\rm s}<10^{14}$ Hz).
There exist low, intermediate, and high synchrotron-peaked BL Lacs (LBLs, IBLs, and HBLs, respectively,
defined by whether $\nu_{\rm s}<10^{14}$ Hz, $10^{14}<\nu_{\rm s}\ (\rm Hz)<10^{15}$, or $\nu_{\rm s}>10^{15}$ Hz) \citep{Abdo2010}.
$\gamma$ rays from HBLs can be produced by SSC,
while $\gamma$ rays from FSRQs are usually attributed to EC.
It is very likely that the jet properties of LBLs are similar to that of FSRQs.

By modeling blazar spectra,
emission mechanisms and physical properties of the relativistic jets can be determined \citep[e.g.,][]{ghisellini14,zhang12}.
Besides, the information of violent variation including variability timescale and profile of light curve also put constraints on the jet properties of blazars.
Although it seems that blazar variability is usually aperiodic, many efforts have been still made to search for the periodic variabilities in blazars
\citep[e.g.,][]{Kidger1992,Bai1998,Bai1999,Fan2000,Xie2008,Li2009,King2013,Urry2011,Gupta2014}.
%which may give us more insights into the physics of blazars.
Naturally, these studies mainly focus on radio and optical-UV band because of the abundant data.

Quasi-periodic oscillations (QPO) have been detected in the X-ray radiation of black hole X-ray binaries (BHXBs) \citep{Re}.
The X-ray QPOs were
found to have two types, low-frequency QPOs and
high-frequency QPOs \citep{Re06}, which are thought to originate in
the inner accretion disk of a black hole.
\citet{Re06} suggested an inverse linear
relation between QPO frequency and black hole mass.
Recently \citet{pan}  and \citet{zhangpf} analyzed the X-ray data of a narrow-line Seyfert 1 galaxy 1H 0707-495, and found a possible QPO.
The  QPO in 1H 0707-495 follows the QPO frequency BH
mass relation suggested by \citet{Re06}.
It is interesting that this relation spans from stellar-mass to supermassive BHs. It may indicate that accretion onto a stellar-mass black
hole is comparable to accretion onto a supermassive black hole.

In this work we focus on QPOs in blazars. 
It should be pointed out that blazar emission is dominated by non-thermal radiation from jet, 
and the X-ray QPOs mentioned above were found in thermal radiation from accretion disk.
Some possible QPO signals were found in radio, optical and X-ray bands.
The period has two types, the short term of several tens days and the long term of several years \citep[e.g.,][]{Rieger}.
QPOs in blazars may give us more insights into the physics of blazars.

Thanks to the Large Area Telescope (LAT) onboard the \emph{Fermi Gamma-ray Space Telescope} \citep{Abdo2009},
running in an all-sky coverage monitoring every $\sim$ 3 hours \citep{Atwood2009,Abdo2009}, we have a long-term view of
the variability of a large sample of $\gamma$-ray blazars.
\emph{Fermi}-LAT has been collecting data for over eight years, allowing us to search for quasi-periodic
variability with a timescale of a few years in gamma-ray flux \citep[e.g.,][]{1553,Sandrinelli2014,Sandrinelli2016a,zhang-yan}.

So far, three blazars (PG 1553+113, PKS 2155-304 and PKS 0537-441) have been reported having quasi-periodic
variability in gamma-ray fluxes with the significance of $\geq3\ \sigma$ \citep{1553,Sandrinelli2014,Sandrinelli2016a,zhang-yan}.
In particular, \cite{zhang-yan} reported a $\gamma$-ray quasi-periodic
variability in PKS 2155-304 with a significance of 4.9 $\sigma$.
The three blazars are HBLs. HBL usually has "clean" environment around the jet, and the $\gamma$ rays
are produced by SSC.
It is interesting to search for quasi-periodic variabilities in other subclasses of blazars, e.g., FSRQs.

In this work we analyze the gamma-ray data from the distant FSRQ PKS 0426-380 in the interval
between 2008 August and 2016 December. We find a significant quasi-periodicity in the $\gamma$-ray flux.
The significance of the 3.35 $\pm$ 0.68 years period is  $\simeq3.6\ \sigma$.
This paper is organized as follows: we describe \emph{Fermi}-LAT data analysis procedures in Section 2;
in Section 3 we show results; summaries and discussions are given in Section 4.

%%%%%
%Section 2 - Observations
%%%%%

\section{Observations and Data reduction}
\label{sec:Observations}

The blazar PKS 0426-380 is classified as a FSRQ \citep{Healey,ghisellini11,Sbarrato}.
PKS 0426-380 has broad lines visible in the low emission
state \citep{Sbarufatti}. PKS 0426-380 has a large $\gamma$-ray luminosity (i.e. $L_{\gamma}>10^{48}\rm \ erg\ s^{-1}$)
and a large ratio between the broad-line region (BLR) luminosity
and the Eddington luminosity (i.e. $L_{\rm BLR}/L_{\rm Edd}\sim10^{-3}$) \citep{ghisellini11}.
PKS 0426-380 is a  distant blazar with the redshift $z=1.11$ \citep{Heidt,Sbarufatti}.

The public $\gamma$-ray data of PKS 0426-380 are obtained from the observations of the LAT.
\emph{Fermi}-LAT is an electron-positron pair production detection sensitive to photon energies from $\sim$ 20 MeV to 300 GeV\footnote{https://fermi.gsfc.nasa.gov/science/instruments/table1-1.html}.
The LAT has a largest effective area $\rm \sim 8000~cm^2$ at 1 GeV, a field of view $\sim$ 2.4 sr,
and a point spread function $< 0.8^{\circ}$ above 1 GeV.

The data are selected between 2008 August 4 and 2016 December 15 (MJD 54682.66-57737.66) with the energy range
from 100 MeV to 500 GeV. The LAT data analysis employ the Fermi Science Tools version v10r0p5 package.
We follow the standard procedure provided in Fermi Science Support Center (FSSC)
\footnote{http://fermi.gsfc.nasa.gov/ssc/data/analysis/scitools/}, to reduce the data.
We select the events in a circular region of interest (ROI) of 15$^\circ$ radius centered on the position of PKS 0426-380.
To minimize the contamination of $\gamma$ rays from the Earth limb, we exclude the events with zenith angle $>~90^{\circ}$.
The diffuse $\gamma$-ray emission of the Galactic and extra-galactic are modeled using the two files:
gll\_iem\_v06.fit and iso\_P8R2\_SOURCE\_V6\_v06.txt. We run the Science Tools \emph{gtmktime} to select the good time intervals.
The SOURCE class of photon-like events of the new Pass 8 data \citep{Atwood2013} is used with the instrument response functions (IRFs) P8R2\_SOURCE\_V6.

A binned maximum likelihood algorithm implemented in \emph{gtlike} is applied between the model file and the events of observation.
The data from a $20^{\circ}\times20^{\circ}$ square ROI are divided into a spatial pixel size of $0.1^{\circ}\times0.1^{\circ}$.
In energy dimensionality, they are binned into 30 logarithmically equal bins. We use the script \emph{make3FGLxml.py} to generate
the model file \footnote{http://fermi.gsfc.nasa.gov/ssc/data/analysis/user/}.
The file is composed of the information (the parameters of spectrum and the spatial positions) of all known 3FGL sources \citep{Acero2015}.
We employ \emph{gtlike} to drive the best-fitting results including flux, photon index, and test statistic (TS) value.
The TS is defined as $\rm TS = -2ln(\mathcal{L}_0/\mathcal{L}_1)$, where $\mathcal{L}_0$ and $\mathcal{L}_1$
are respectively the maximum likelihood value of
the model with and without the source, and it describes the significance of source.
We save the best-fitting results into a new model file. The following analysis of light curves are base on this new model file.
The results in this paper are only given with statistical errors.

\section{Results}

\subsection{$\gamma$-ray light curves}
\label{subsec:make lc}

\emph{Fermi}-LAT collaboration provides two techniques to generate light curves,
i.e. the maximum likelihood optimization and the exposure-weighted aperture photometry \citep{Corbet2007,Kerr2011}.

\begin{figure}
\centering
	\includegraphics[width=260pt,height=230pt]{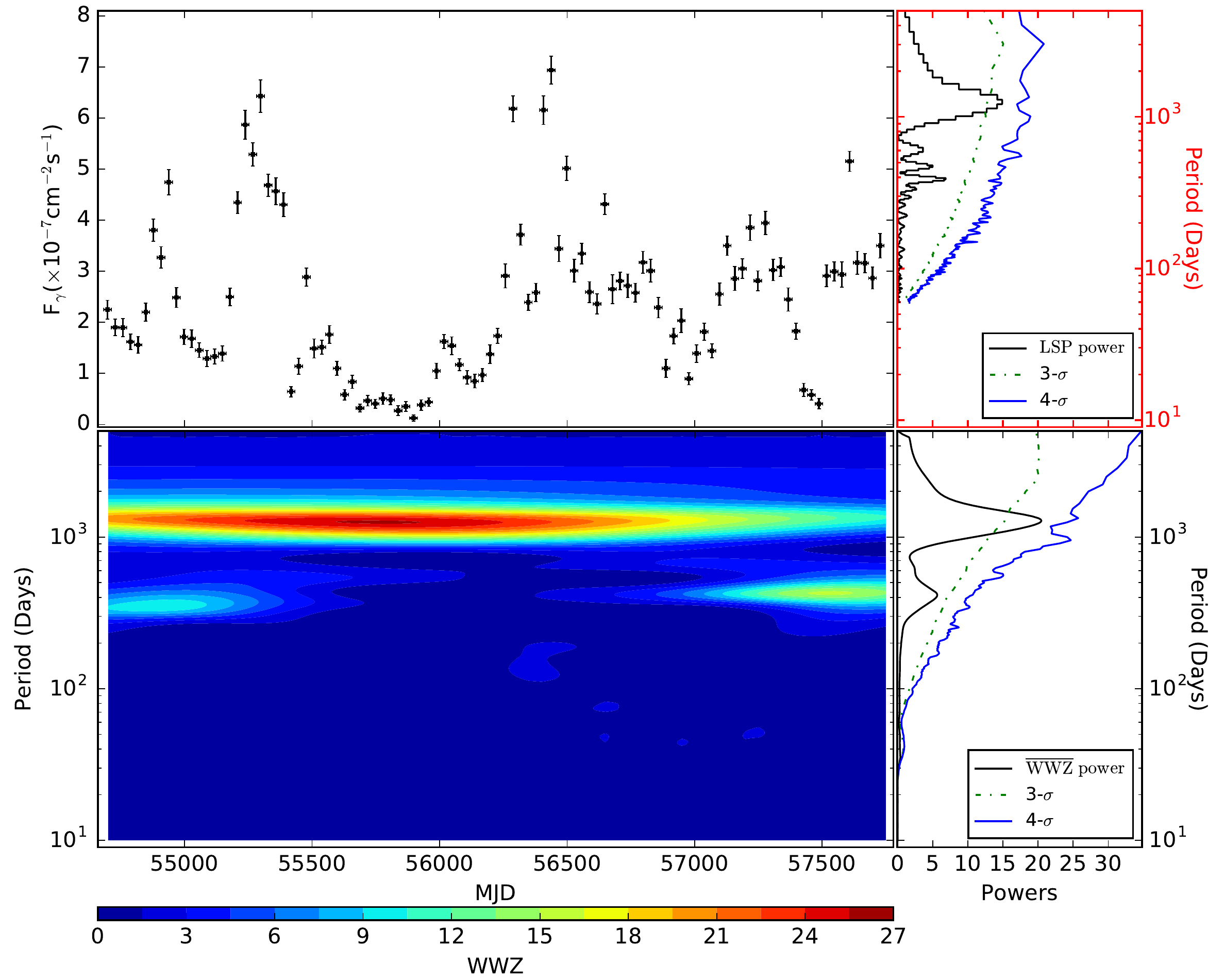}
		\caption{Upper left panel: the likelihood gamma-ray light curve of PKS 0426-380 above 100 MeV with the one-month bin.
		         Upper right panel: the corresponding LSP power spectrum; the blue solid line and
		                                       green dashed line represent the 4 $\sigma$ and 3 $\sigma$ confidence level, respectively.
		         Lower left panel: the 2D plane contour plot of the WWZ power of monthly light curve.
		         Lower right panel: the black solid line represents the time-averaged WWZ power;
		                            the blue solid line and green dashed line represent the 4 $\sigma$ and 3 $\sigma$ confidence level,
		                            respectively.}
	\label{lc_m}
\end{figure}

\begin{figure}
\centering
	\includegraphics[width=260pt,height=230pt]{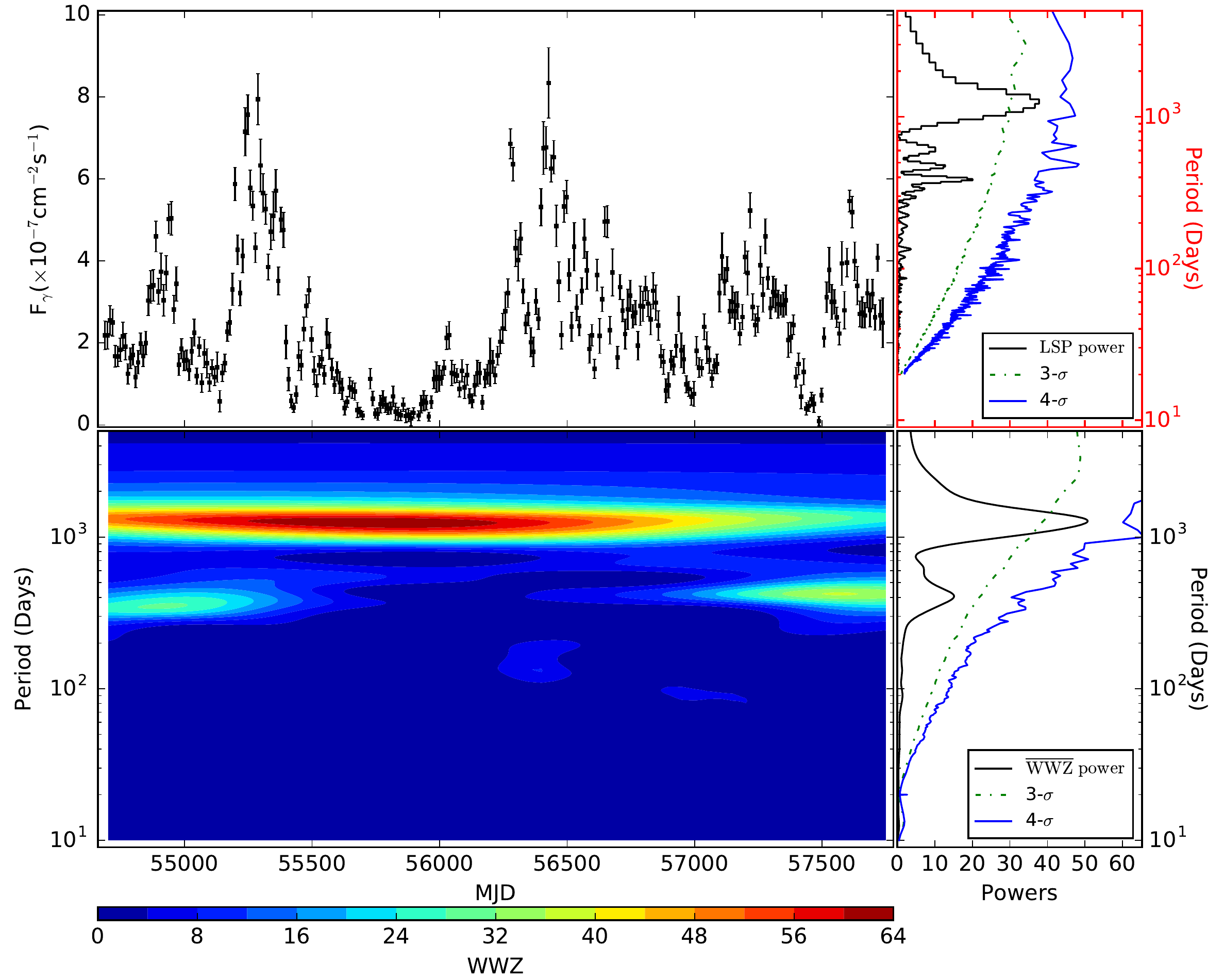}
		\caption{Same as Fig.~\ref{lc_m} but with the 10-day bin.}
	\label{lc_10days}
\end{figure}

In the analysis with the maximum likelihood optimization, the light curves are first built by using a one-month bin.
We employ a unbinned maximum likelihood fitting technique and run the ScienceTools \emph{gtlike}
to obtain the flux and TS value for each time bin.  In this step, the events are selected in a circular ROI of
15$^\circ$ radius centered on the coordinate of the target.
The model is the same as the new model file mentioned in Section 2, except the parameters of the spectrum shape are frozen.
The light curve is shown in Fig. \ref{lc_m}. To test the impacts of different time bins on the results,
we also build the light curve with a 10-day bin over 0.1 GeV (Fig. \ref{lc_10days}).

%%%%

\begin{figure}
\centering
	\includegraphics[width=260pt,height=230pt]{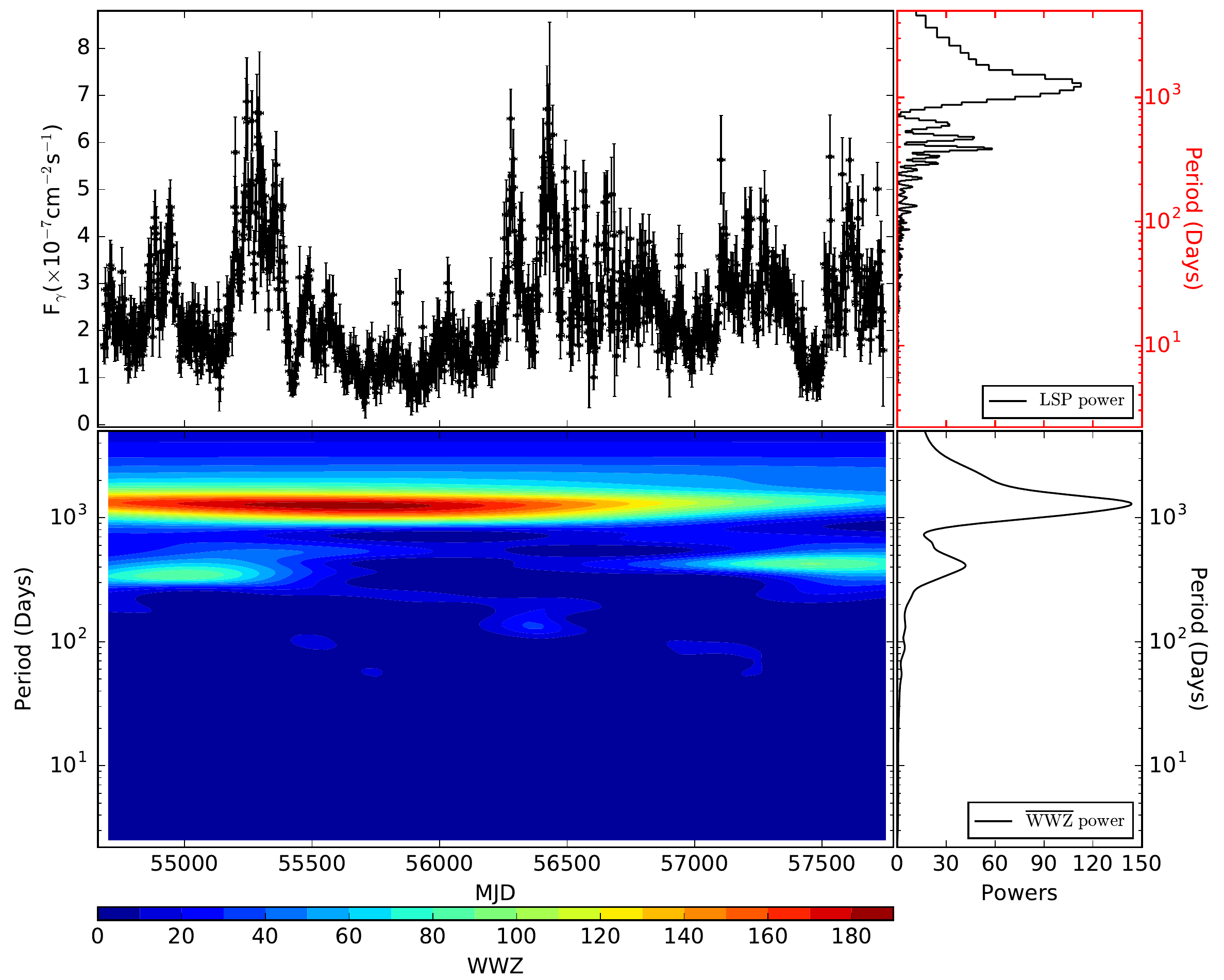}
		\caption{Upper left panel: the weighted-photon light curve of PKS 0426-380 above 100 MeV with the 2.5-day bin.
		         Upper right panel: the black solid line represents the LSP power spectrum.
		         Lower left panel: the 2D plane contour plot of the WWZ power of 2.5-day light curve.
		         Lower right panel: the solid line represents the time-averaged WWZ power.}
	\label{ap}
\end{figure}

In the analysis with the exposure-weighted aperture photometry,
we generate light curve with a modified version of aperture photometry.
We exclude the events during the period when the target is within 5$^\circ$ of the Sun.
The probability of each photon from a specific source is calculated by the ScienceTools \emph{gtpsrcrob}
with the best-fitting model and IRFs P8R2\_SOURCE\_V6. Then we sum the probability of the events within 3$^\circ$ radius circle
region centered on the position of PKS 0426-380 over 100 MeV. The 2.5-day bin is used.
Then the exposure of each time bin is determined with the ScienceTools \emph{gtexposure}.
The light curve weighted by the exposure of each time bin is shown in Fig. \ref{ap}.

\subsection{Searching for $\gamma$-ray quasi-periodic modulation}

The Lomb-Scargle Periodogram \citep{Lomb1976,Scarle1982} is a widely used tool for searching for period of variability. 
However, this method does not take into account the fluctuations of periodic signal with time.
Therefore, we also use the WWZ technique with a Morlet mother function \citep{Foster1996} for testing the result of LSP.
In Figs.~\ref{lc_m}-\ref{ap} we also show the corresponding LSP and WWZ powers. 
%the 2D plane contour plot of the WWZ power is shown in the lower left panel of Fig. \ref{lc_m}
%and the time-averaged WWZ power is shown in the lower right panel of Fig. \ref{lc_m};
%We also test the impacts of different time-length and energy range on the signal.
%Using the same techniques, we analyze the the light curves of 10-day bins with the method of maximum likelihood optimization,
%30-day in energy range from 1.0 to 50 GeV with same method as former,
%and 2.5-day bin with the method of exposure-weighted aperture photometry for the construction of the LAT light curve.
%The results of this three light curves confirm the results of 1-month light curve (see the Figs. \ref{lc_10days}-\ref{lk_50GeV}).
All the powers show a obvious peak at $\sim$ 1222 days (i.e., 3.35 $\pm$ 0.68 years).
The uncertainty of the signal is evaluated based on the half width at half maximum (HWHM)
of the Gaussian fitting at the position of the power peak.
We first apply a simple method for testing the significance of peaks in the periodogram of red noise data
\citep{vau} to all the highest power-peaks in Figs.~\ref{lc_m}-\ref{ap},  
and derive a similar significance of $\sim3.5\ \sigma$.
For obtaining the precise significance of the signal, we should simulate light curves based on the obtained best-fitting result of power spectral density (PSD) and the probability density function (PDF) of observed variation \citep[e.g.,][]{zhang-yan}. The details of the simulation and significance estimation are provided by \citet{Emmanoulopoulos2013}, \citet{1553} and \citet{Bhatta2016}.
Following the procedure,
we simulate $3\times10^{4}$ light curves with the DELCgen program, 
and then we evaluate the significance of the signal. The significance of the signal is $\simeq3.6~\sigma$.

\begin{figure}
\centering
	\includegraphics[width=260pt,height=230pt]{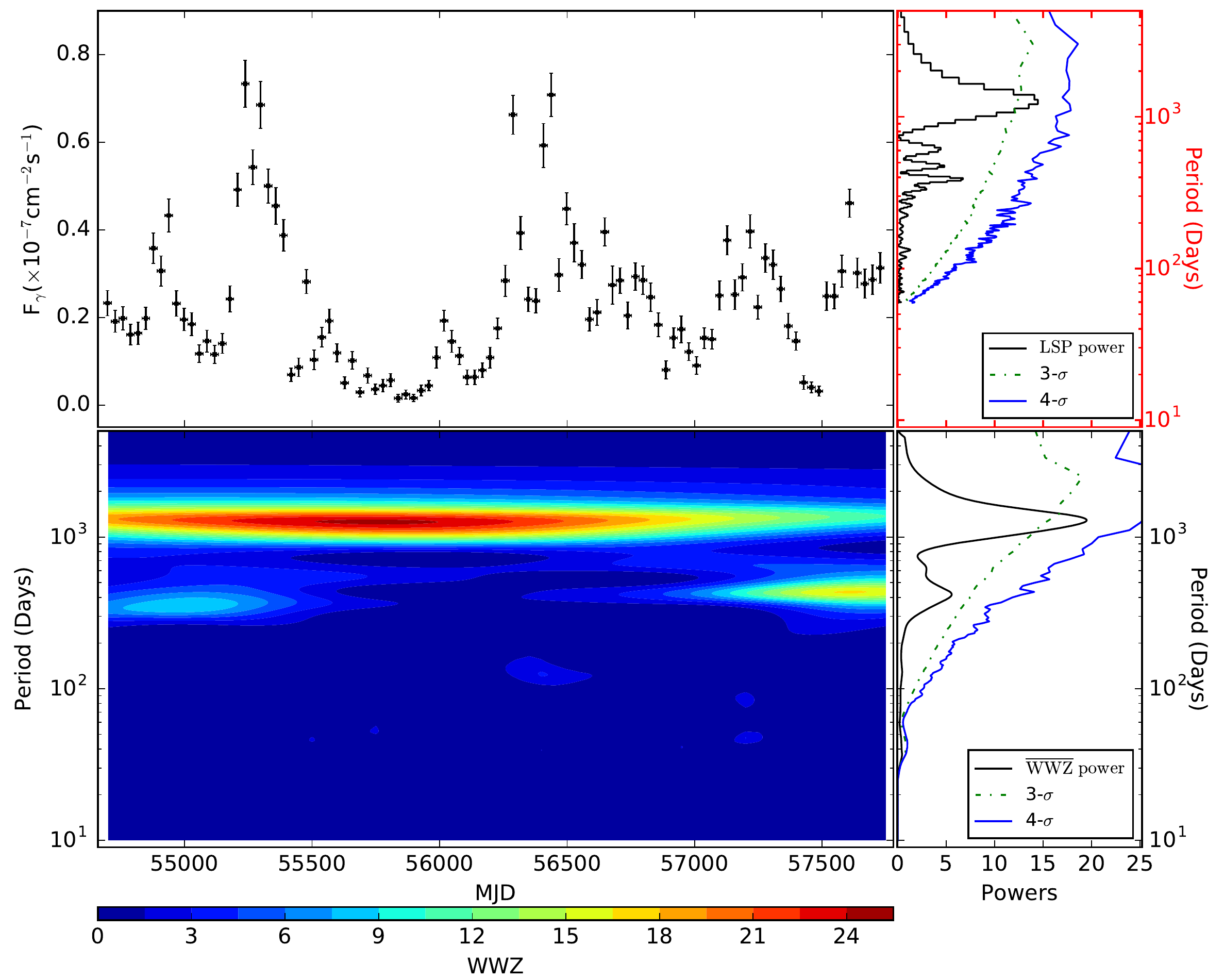}
		\caption{Same as Fig.~\ref{lc_m} but with the energy range of 1-50 GeV.}
	\label{lk_50GeV}
\end{figure}

Photons above 100 GeV from PKS 0426-380 are found in our analysis \citep[also see][]{Tanaka}.
In order to avoid the effect of the absorption by extragalactic background light (EBL),
We build the light curves in the energy range of 1-50 GeV (Fig.~\ref{lk_50GeV}).
Again we find a strong signal near the period of $\sim$ 1222 days in the corresponding LSP and WWZ powers.

\begin{figure}
\centering
	\includegraphics[width=260pt,height=140pt]{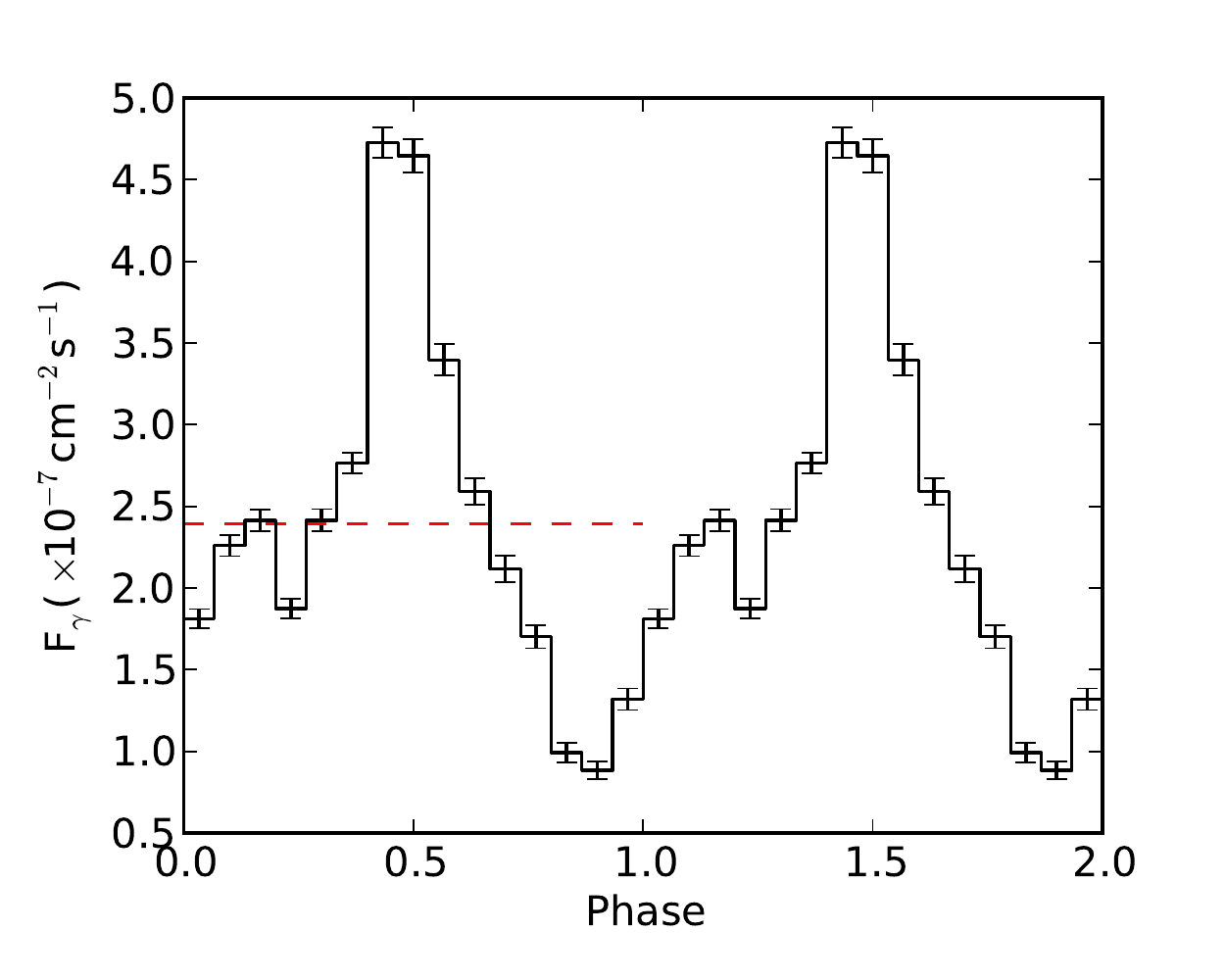}
		\caption{The epoch-folded pulse shape above 0.1 GeV with a period of 1221.9 days;
		         the red dashed line is the mean flux; for clarity, we show two period cycles.}
	\label{fold_lc}
\end{figure}

Using phase-resolved likelihood analysis method, we fold the gamma-ray photons from the 20$^\circ\times20^\circ$
square region centered on the position of the target. We calculate a phase for every event in the region
according to the period and the arrival time (with phase zero corresponding to MJD 54682.66). The pulse interval is divided into 15 segments.
For each segment, we use a binned likelihood technique to obtain the flux with the same model file
as generating light curve. The folded light curve is shown in Fig. \ref{fold_lc}.
We fit the folded light curve with a constant flux, and derive the reduced $\chi^2_{\rm min}=3156.5/14$.
The null hypothesis that the phase-resolved light curve is steady can be rejected at a far exceeding 5-$\sigma$ confidence-level.
The folded light curve in Fig. \ref{fold_lc} also confirms the signal mentioned above, and indicates that the gamma-ray flux varies with phase.
%%%%%%%%%%%%%%%%%%%%%%%%%%%%%%%%%%
%%%%%%%%%%%%%%%%%%%%%%%%%%%%%%%%%%%%%%%%%%%%%
\subsection{Searching for optical-UV quasi-periodic modulation}
\label{subsec:qpo}

\begin{figure}
\centering
	\includegraphics[width=260pt,height=230pt]{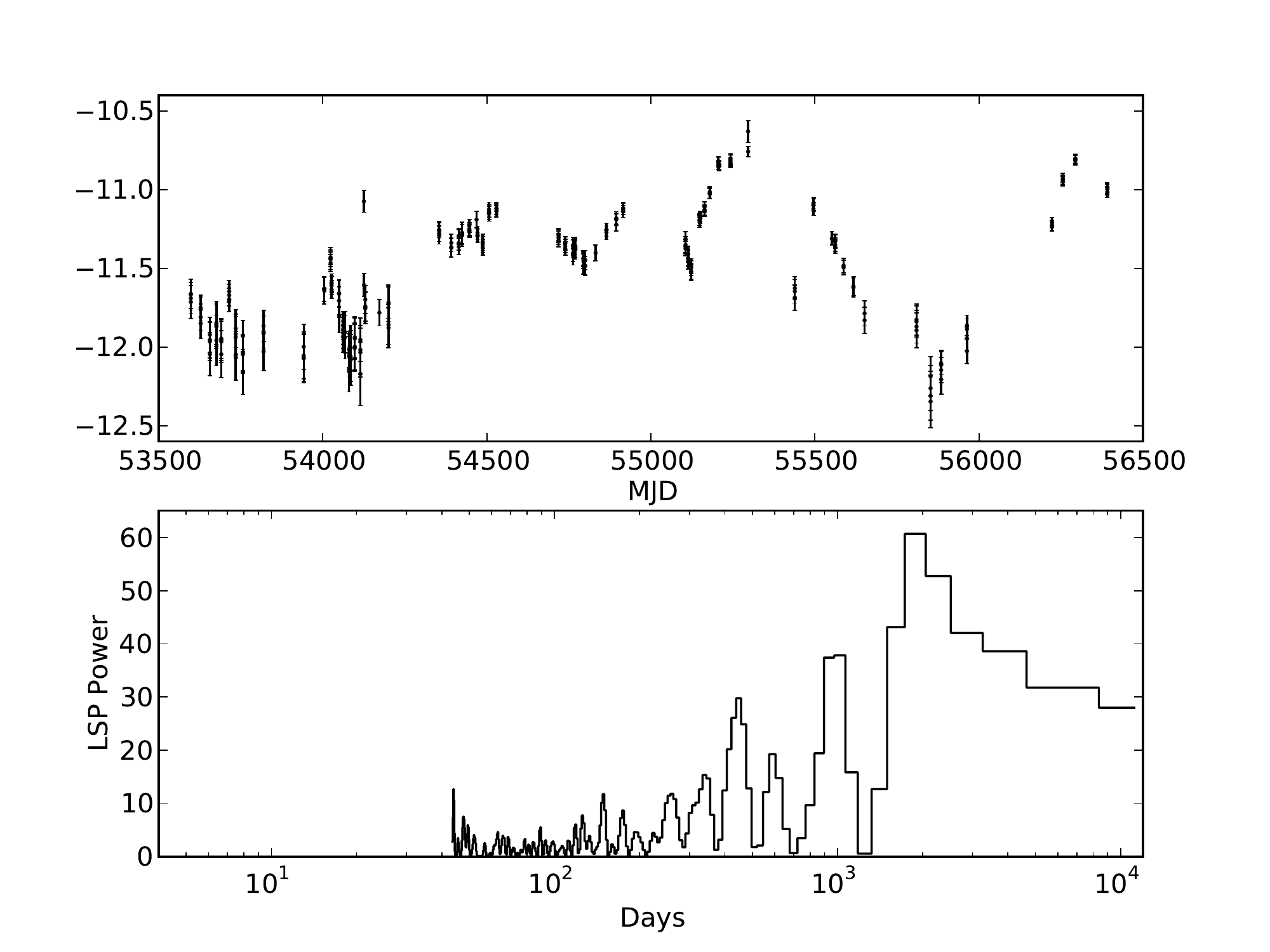}
		\caption{Upper panel: the light curve of optical-UV data for PKS 0426-380 (data from ASI SCIENCE DATA CENTER).
		         Lower panel: the corresponding LSP power spectrum.}
	\label{ur}
\end{figure}

We have tried to collect long-term multiwavelength archive data of PKS 0426-380.
We obtain the optical-UV data covering from 2005 August to 2013 April provided by ASI SCIENCE DATA CENTER (see Fig.~\ref{ur}).
The optical-UV flux is analyzed using the
same techniques as for the $\gamma$-ray flux but does
not show any significant periodic signal.
%%%%%
%Section 3 - SUMMARY
%%%%%

\section{SUMMARY AND DISCUSSION}
\label{sec:summary}

We have produced the $\gamma$-ray light curves of a distant FSRQ PKS 0426-380 with the latest LAT data.
We have found a significant quasi-periodic variability in the $\gamma$-ray flux with a period of 3.35 $\pm$ 0.68 years and a significance of $3.6\ \sigma$. 
More multiwavelength data are needed to confirm the periodicity in PKS 0426-380.
If the $\gamma$-ray quasi-periodic variability is true, the next $\gamma$-ray flux peak will occur in 2019.

The archive optical-UV data covering from 2005 August to 2013 April have been also analyzed but no quasi-periodic signal is found. 
This may be due to the incomplete of the data.
On the other band, optical-UV emission from FSRQs may be dominated by accretion disk photons.
This complicates the confirmations of $\gamma$-ray quasi-periodic variability in FSRQs.

$\gamma$-ray quasi-periodic modulations with the significance of $\geq3\ \sigma$ have been reported
in HBLs \citep{1553,Sandrinelli2014,Sandrinelli2016a,zhang-yan}.
PKS 0426-380 is the first FSRQ having a significant $\gamma$-ray quasi-periodic variability.
Although PKS 0426-380 has a longer observed period ($T_{\rm obs}\sim3$ year) than the three HBLs ($T_{\rm obs}\sim2$ year), 
the intrinsic periods ($T_{\rm int}$) for PKS 0426-380 and the three HBLs are same because of $T_{\rm obs}=T_{\rm int}(1+z)$.
It is well known that FSRQs have larger black hole mass and higher accretion rate compared to HBLs \citep[e.g.,][]{ghisellini11,ghisellini14}.
It is interesting that FSRQs and HBLs have same period.
It may indicate that the $\gamma$-ray quasi-periodic modulations in blazars are not solely linked to the central black hole and the accretion disk.

A possible interpretation for the $\gamma$-ray QPOs is a 
varying Doppler boosting for a periodically changing
viewing angle, which is related to a helical jet \citep[e.g,][]{Rieger04,Ko2016}.
The helical structure in radio jet of AGN has been detected \citep[e.g,][]{Rieger}.
One way to explain the helical jet is involving the presence
of a binary supermassive black hole (SMBH) \citep[e.g,][]{Ko2016,Sobacchi}.
In the scenario of binary SMBH, one can estimate the parameters of the binary system, e.g.,  the separation between the two SMBHs \citep[e.g.,][]{Sobacchi}, 
$$D\sim10^{16}\frac{1+q}{q}\frac{T_{\rm int}}{2\ \rm yr}\ \rm cm\ ,$$
where $q$ is the mass ratio of the
secondary SMBH to the primary SMBH. Assuming $q\sim1$ and using $T_{\rm int}\sim 2\rm yr$, we obtain $D\sim 0.003\ $pc.
The two SMBHs will merge due to emission of gravitational waves. The time-scale for SMBHs merging  is \citep[e.g.,][]{Sobacchi}, 
$$t_{\rm m}\sim4\times10^{4}q(\frac{q}{1+q})^3\frac{T_{\rm int}}{2\ \rm yr}\ \rm yr\ .$$
Using $q\sim1$ and $T_{\rm int}\sim 2\rm yr$, we obtain $t_{\rm m}\sim 4000 \ $yr.

It should be kept in mind that the binary SMBH scenario is not the unique explanation for the $\gamma$-ray QPOs.
Alternative models,  for instance disk oscillations or
Kelvin-Helmholtz instabilities, have been proposed \citep[see][for a review]{Ko2016}. 
Nevertheless we suggest that a cumulative multifrequences blazar QPOs can be used to probe the underlying physical mechanism, and may also shed light on the questions of galaxies merger. Moreover, multifrequences QPOs may put new constraint on blazar emission models \citep{zhang-yan}. 

\section*{ACKNOWLEDGEMENTS}
We thank the referees for constructive suggestions that
significantly improved the manuscript. Part of this work is based on archival data, software or online services provided by the ASI SCIENCE DATA CENTER (ASDC).
We acknowledge the financial support from the 973 Program of China under grant 2013CB837000, the National Natural Science Foundation of China (NSFC-11573060 and NSFC-11573026), Key Laboratory of Astroparticle Physics of Yunnan Province (No. 2016DG006). The work of Dahai Yan is supported by CAS ``Light of West China" Program.

\appendix
\section{Autocorrelation analysis and $\gamma$-ray QPO in blazar}

\begin{figure}
\centering
	\includegraphics[width=260pt,height=200pt]{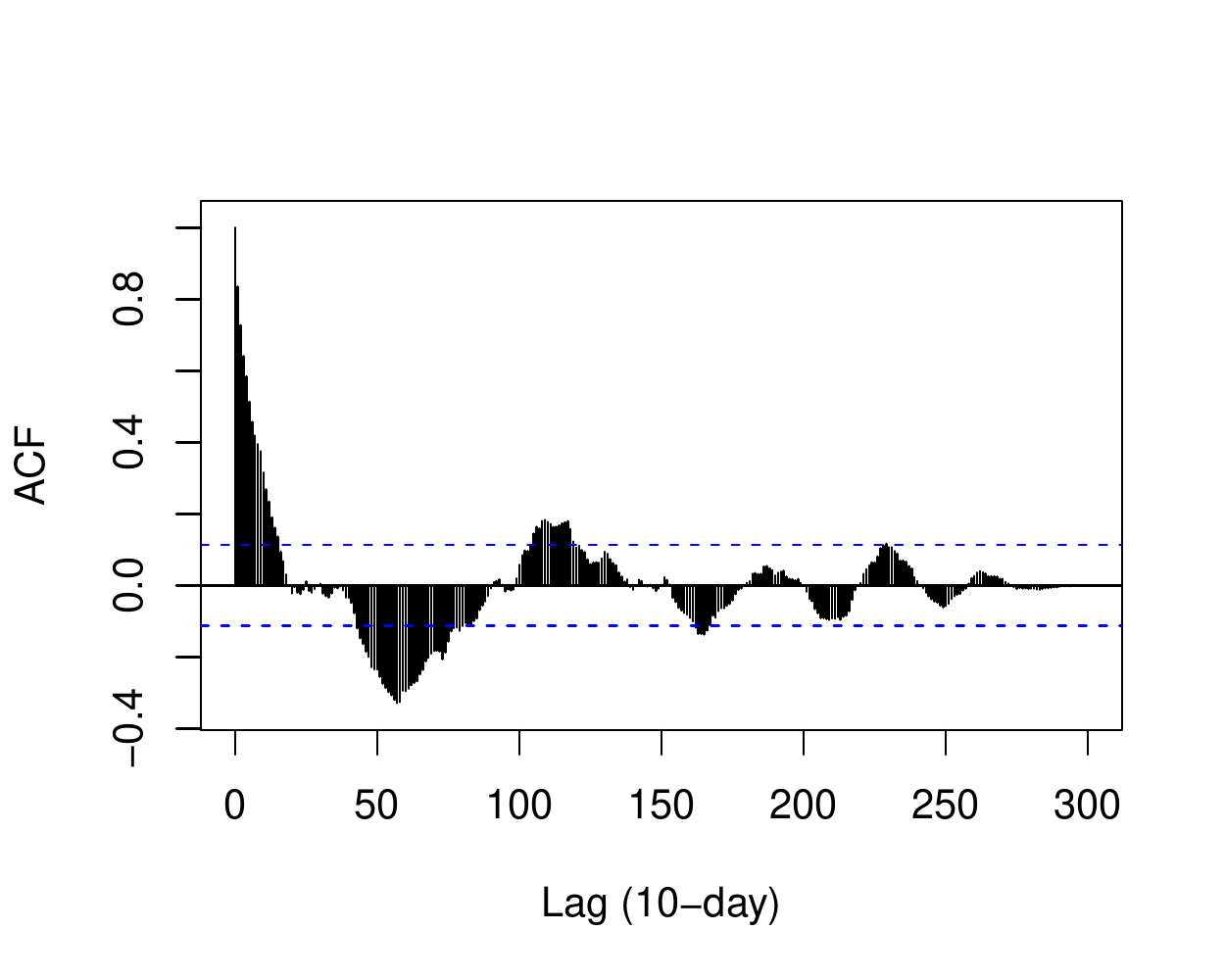}
		\caption{ACF of the 10-day bin likelihood gamma-ray light curve. The dashed horizontal lines represent the 95\% confidence level.}
	\label{arima}
\end{figure}

In addition to the methods  based on Fourier analysis (e.g., LSP and WWZ), 
an autocorrelation analysis also has been used in light curve analysis to search for quasi-periodic behavior
or variations with a characteristic timescale \citep[e.g.,][]{Li}.
A regularly periodic
light curve will have strong correlations at time lags that are
multiples of the period. For an
evenly spaced time series, the autocorrelation function (ACF) is the
fraction of the total variance due to correlated values at time lags.
We apply an autocorrelation analysis based on the
algorithm of \citet{box} to the $\gamma$-ray data.
We calculate the ACF with the parametric autoregressive integrated moving average (ARIMA) model
\citep[e.g.,][]{Hamilton1994,Chatfield2003}.
In Fig.~\ref{arima}, we show the ACF of the 10-day bin likelihood gamma-ray light curve.
It can be seen that two positive correlation peaks appear respectively at the lag time of $\sim$1200 day and $\sim$2300 day.
The result of autocorrelation analysis is consistent with the result of Fourier analysis.

%%%%%%
% Bibliography %
%%%%%%

\bibliography{ApJ}

\end{document}